\documentclass{elsart}

 \usepackage{epsfig}

\usepackage{amssymb}

\begin{document}

\begin{frontmatter}



\title{Suppression of heavy flavors at RHIC \& LHC}


\author{Carlos A. Salgado\thanksref{label1}}
\address{Dipartimento di Fisica, Universit\`a  di Roma ``La Sapienza'' \\
 and INFN, Roma, Italy.}

\thanks[label1]{Permanent address: Departamento de F\'\i sica de Part\'\i culas, Universidade de Santiago de Compostela (Spain).}

\begin{abstract}
Some of the open questions on {\it jet quenching} are expected to be clarified by measuring heavy--flavored mesons at high transverse momentum. The formalism based on radiative in-medium energy--loss, which describes other high-$p_t$ results at RHIC, gives definite predictions for the suppression of charm and beauty quarks. However, the uncertainties from both contributions to the observed electrons make the interpretation of the data difficult due to the absence of a well calibrated benchmark. We review the present situation as well as the consequences for the future LHC. We also comment on the use of heavy flavored jet angular correlations as an additional tool to study the underline dynamics of jet quenching. 
\end{abstract}

%\begin{keyword}
%heavy--flavor production \sep jet quenching \sep heavy--ion collisions
%\PACS 
%\end{keyword}
\end{frontmatter}

% main text
\section{Introduction}
\label{}

Are non-photonic electrons data measured in AuAu collisions at RHIC \cite{Adler:2005xv,Abelev:2006db} a chalenge for the {\it radiative energy loss} formalism \cite{Baier:1996sk,Salgado:2003gb}? The new data present a very strong suppression, similar to the one reported for light mesons \cite{Adams:2005dq}. In general terms, the formalism predicts a hierarchy of larger energy loss for gluons than for light quarks and this larger than for heavy quarks. (Strictly speaking this is not a particular feature of the radiated energy loss, but a rather general result from color and mass factors.) The apparent discrepancy is sometimes interpreted as a failure of the formalism and lead to a renewed interest for the so-called {\it collisional energy loss} \cite{collisional} as a possible additional source of the heavy quark energy degradation in the medium. To address this question I will only concentrate on the radiative contribution, referring the interested readers to the relevant contributions in these proceedings for additional discussion \cite{collisional}. 

The original proposal to use heavy quarks as additional constrains to the radiative in-medium energy loss \cite{Dokshitzer:2001zm} 
made an analogy with the vacuum case and propose a suppression of the induced gluon radiation by a {\it dead cone} multiplicative factor. Although most physics relevant effects were correctly identified in \cite{Dokshitzer:2001zm}, a full calculation of mass effects revealed a richer structure \cite{Armesto:2003jh,Zhang:2003wk,Djordjevic:2003zk} and, in fact, a smaller energy loss effect \cite{Armesto:2003jh} than the one reported in \cite{Dokshitzer:2001zm}. In the next sections we will present the formalism used to predict the effects on heavy flavor suppression in heavy--ion collisions and how the benchmark uncertainties in the perturbative cross section, if taken seriously, make unclear the comparison of the available formalism with experimental data. We finish with a proposal to study heavy--quark jets angular correlations.

\section{The jet quenching formalism}

Jet quenching belongs to a larger class of probes of the high--density matter created in heavy--ion collisions characterized for the large virtualities involved, which allow a perturbative treatment. In the typical hard cross section
\begin{equation}
\sigma^{AB\to h}=
f_A(x_1,Q^2)\otimes f_B(x_2,Q^2)\otimes \sigma(x_1,x_2,Q^2)\otimes D_{i\to h}
(z,Q^2)\, ,
\label{eqhard}
\end{equation}
the medium fragmentation function is modeled as \cite{Wang:1996yh}
\begin{equation}
D_{i\to h}^{\rm med}(z,Q^2)=P_E(\epsilon)\otimes D_{i\to h}(z,Q^2).
\label{eqff}
\end{equation}
Here, the {\it quenching weights} $P_E(\epsilon)$ give the probability of an additional in-medium energy loss and is normally computed by assuming a simple Poisson distribution with the medium-induced gluon radiation as input \cite{Baier:2001yt,Salgado:2003gb}. 
\begin{equation}
  P_E(\epsilon) = \sum_{n=0}^\infty \frac{1}{n!}
  \left[ \prod_{i=1}^n \int d\omega_i \frac{dI^{\rm med}(\omega_i)}{d\omega}
    \right]
    \delta\left(\epsilon -\sum_{i=1}^n {\omega_i \over E} \right)
    \exp\left[ - \int d\omega \frac{dI^{\rm med}}{d\omega}\right].
  \label{eqqw}
\end{equation}
Once the geometry of the system is correctly taken into account, a fit to RHIC light--meson data gives the value of the time-averaged transport coefficient $\hat q\sim 5...15$ GeV$^2$/fm \cite{Eskola:2004cr}. This large value and the corresponding uncertainty is a direct consequence of the surface trigger bias effect in inclusive particle suppression measurements \cite{Muller:2002fa,Eskola:2004cr}. This is an intrinsic limitation of inclusive measurements for the characterization of the medium and for the study of the dynamics underlying the propagation of highly energetic partons through a dense medium. Further constraints can be found by i) measuring different particles species, and in particular heavy quarks, as the formalism predicts the hierarchy $\Delta E_g>\Delta E_q^{\rm m=0}>\Delta E_Q^{\rm m\neq 0}$; ii) by directly measuring the induced radiation, i.e. by reconstructing the jet structure in a heavy ion collision.

\begin{figure}
\begin{center}
\includegraphics[width=0.7\textwidth]{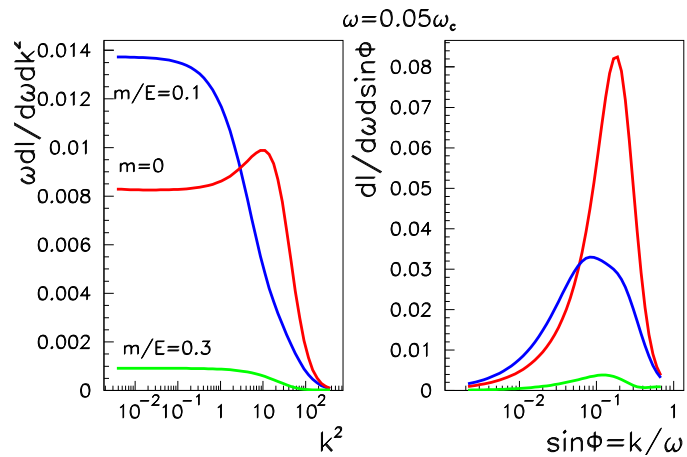}
\end{center}
\caption{$k_t^2$ (left) and angular (right) spectra of medium--induced radiated gluons off a massless (red), charm (blue) and bottom (green) quarks \cite{Armesto:2003jh}.}
\label{fig:spec}
\end{figure}

The needed ingredient to generalize eq. (\ref{eqqw}) to the massive case is the spectrum of radiated gluons $dI^{\rm med}/d\omega$, which has been computed in \cite{Armesto:2003jh,Zhang:2003wk,Djordjevic:2003zk}. Important qualitative properties of these spectra can be simply understood by formation time effects. When the formation time of the radiated gluon becomes larger than the length of the medium a saturation phenomena appears and the spectrum is suppressed with respect to the incoherent case. More specifically, for a massless particle $t_{\rm form}\sim \omega/k_t^2$, and this happens for small [large] values of $k_t$ [$\omega$]. As a result, even for the massless case, the spectrum is not collinear divergent in contrast with the vacuum case -- see Fig. \ref{fig:spec}. In the massive case two opposite effects appear, on the one hand, the radiation is suppressed due to mass terms in the propagators, in complete analogy with the vacuum; on the other hand, the formation time of a gluon radiated from a heavy quark is shorter and the suppression due to formation time effects smaller. While the dominant effect is the first and, hence, the energy loss is smaller, in limited regions of phase space the small angle spectrum can be enhanced as seen in Fig. \ref{fig:spec} \cite{Armesto:2003jh}.

\section{The uncertainties in the benchmark cross sections}

\begin{figure}
\begin{center}
\includegraphics[width=0.5\textwidth,angle=-90]{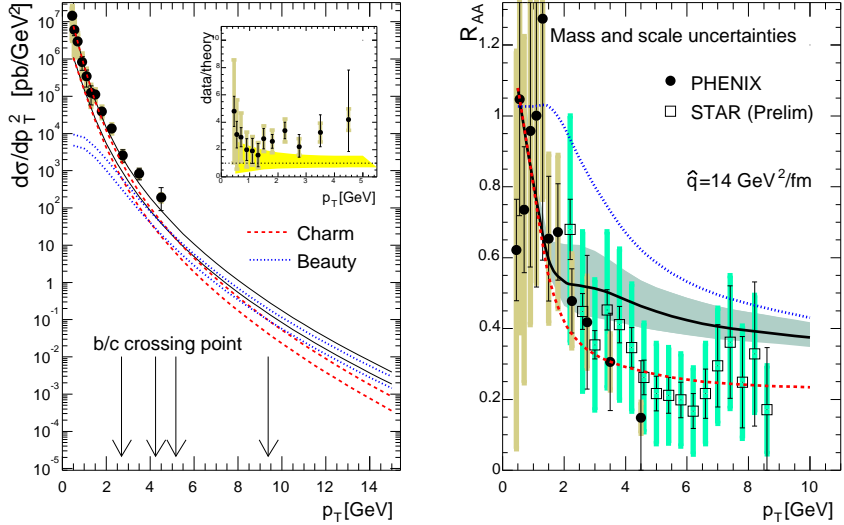}
\end{center}
\caption{Left: Comparison of the FONLL calculation of single 
inclusive electrons from pp collisions at $\sqrt{s}$=200GeV \protect\cite{Cacciari:2005rk}. Right: The nuclear modification factor of electrons with the corresponding uncertainty coming from the perturbative benchmark on the relative $b/c$ contribution. Figure from \protect\cite{Armesto:2005mz}; data from \protect\cite{Adler:2005xv,Abelev:2006db}.}
\label{fighq}
\end{figure}

One advantage of the RHIC program is the possibility of colliding different systems (pp, dA and AA) at the same energy. This provides accurate benchmarks on top of which the effects due to the presence of a medium in AA collisions can be calibrated. This has been the case in particular for the high-$p_t$ suppression of light mesons, for which the importance of a control dAu run was essential \cite{Adams:2005dq}.  Present data on heavy--quark energy loss lacks, however, of such a well calibrated benchmark: a disagreement persists between the proton--proton data measured in pp collisions at $\sqrt{s}$=200 GeV at RHIC and the state-of-the-art FONLL calculation \cite{Cacciari:2005rk} -- see Fig. \ref{fighq}. The experimental situation is still not clear, PHENIX results \cite{Adare:2006hc} are in better agreement with FONLL -- data is a factor of 2 above the central theory value but within the uncertainty band -- than STAR, which report a factor of 5 disagreement \cite{Abelev:2006db}.
The disagreements between heavy flavor hadroproduction data and theory is not something new \cite{Cacciari:2004ur,Mangano:2004xr}, but improvements in theory and experiment led in the past to a rather good description of bottom \cite{Mangano:2004xr} as well as charm data \cite{Cacciari:2003zu}, with, possibly, some tendency to be underestimated by theory for the last. 

Other sources, different from the decay of heavy quarks, could contribute to the yield of observed electrons. In particular, at high enough electron transverse momentum, the Drell-Yan contribution would dominate. It has been shown in \cite{Armesto:2005mz} that this contribution is small even for the largest available values of $p_t$ measured at present at RHIC.

\section{How uncertainties in the benchmark translate into $R_{AA}$}

The mass effects in the radiative energy loss formalism depend on the ratio $M/E$ -- mass over energy of the quark. The mass of the charm quark is found to be not large enough for these effects to produce a strong effect and the predicted suppression for charm quarks is similar to that of light quarks \cite{Armesto:2005iq}. For bottom, however, the effects are found to be much larger. One should take into account that due to kinematic effects, the correlation between the measured electron $p_t$ and that of the original quark is rather weak and, in fact, the most relevant region for present data is that of large heavy quark transverse momenta.

Once the transport coefficient of the medium has been fixed by other set of data, the prediction for the jet quenching effects in heavy quarks is parameterless. In Fig. \ref{fighq} we plot the corresponding $R_{AA}$ ratios for electrons coming from charm and bottom decays for $\hat q=$ 14 GeV$^2$/fm. The prediction for the $R_{AA}$ of the total non-photonic electrons needs, however, a good control on the relative contributions of both flavors. Within the perturbative approach, the uncertainties are estimated by changing the mass of the heavy quarks and the renormalization, factorization and fragmentation scales -- see \cite{Cacciari:2005rk} for details. This procedure leads to a factor of $\sim$ 2 uncertainty for both flavors and the corresponding band for the proton-proton predictions. Applying this same procedure, we obtain the uncertainty band for the $R_{AA}$ case plotted in Fig. \ref{fighq}, i.e. this band is just the reflection of changing the masses of the heavy quarks and the scales within reasonable limits (the same as in proton-proton) in the perturbative cross section. We find the description of the data by our jet quenching formalism reasonable within the uncertainties due to the perturbative benchmark, specially before the disagreement in proton--proton data is understood. Notice that a larger contribution from the charm quark would translate into a better agreement with the large suppression reported.

\section{The lessons for the LHC}

\begin{figure}
\begin{center}
\includegraphics[width=0.7\textwidth]{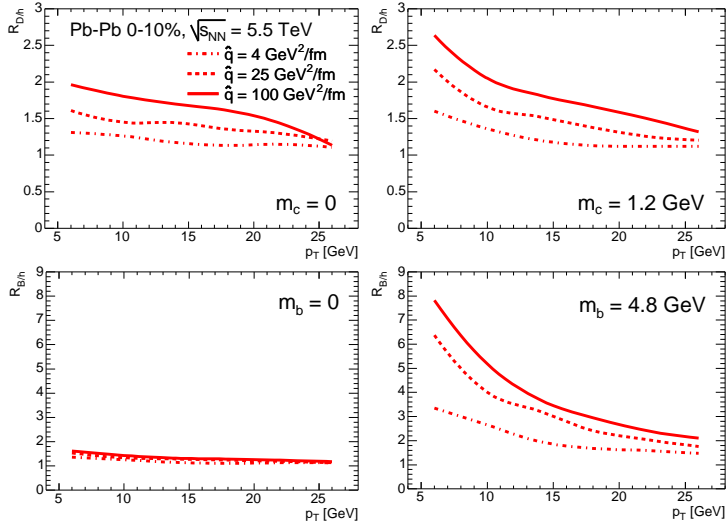}
\end{center}
\caption{Heavy-to-light ratios for $D$ mesons (upper plots) and $B$ mesons (lower
plots) for the case of a realistic heavy quark mass (plots on the right) and for a case study in which the quark mass dependence of parton energy loss is neglected (plots on the left).}
\label{figlhc}
\end{figure}

At the LHC a direct measurement of heavy mesons is under study by the different experimental collaborations and seems feasible \cite{andrea}. As shown in the previous section, this is an important prerequisite for the study of the underline dynamics of jet quenching with heavy quarks. Concerning the inclusive measurements, the charm quark is again expected to be suppressed as much as the light quarks, while the larger bottom quark mass will allow a study of mass effects in energy loss. The strategy \cite{Armesto:2005iq} is to measure color charge dependence of the energy loss by charm-to-light ratios and the mass dependence by bottom-to-light ratios -- see Fig. \ref{figlhc}.

\section{Heavy quark jet shapes?}

\label{sec:evol}

\begin{figure}
\begin{minipage}{0.5\textwidth}
\begin{center}
\includegraphics[width=0.75\textwidth,angle=-90]{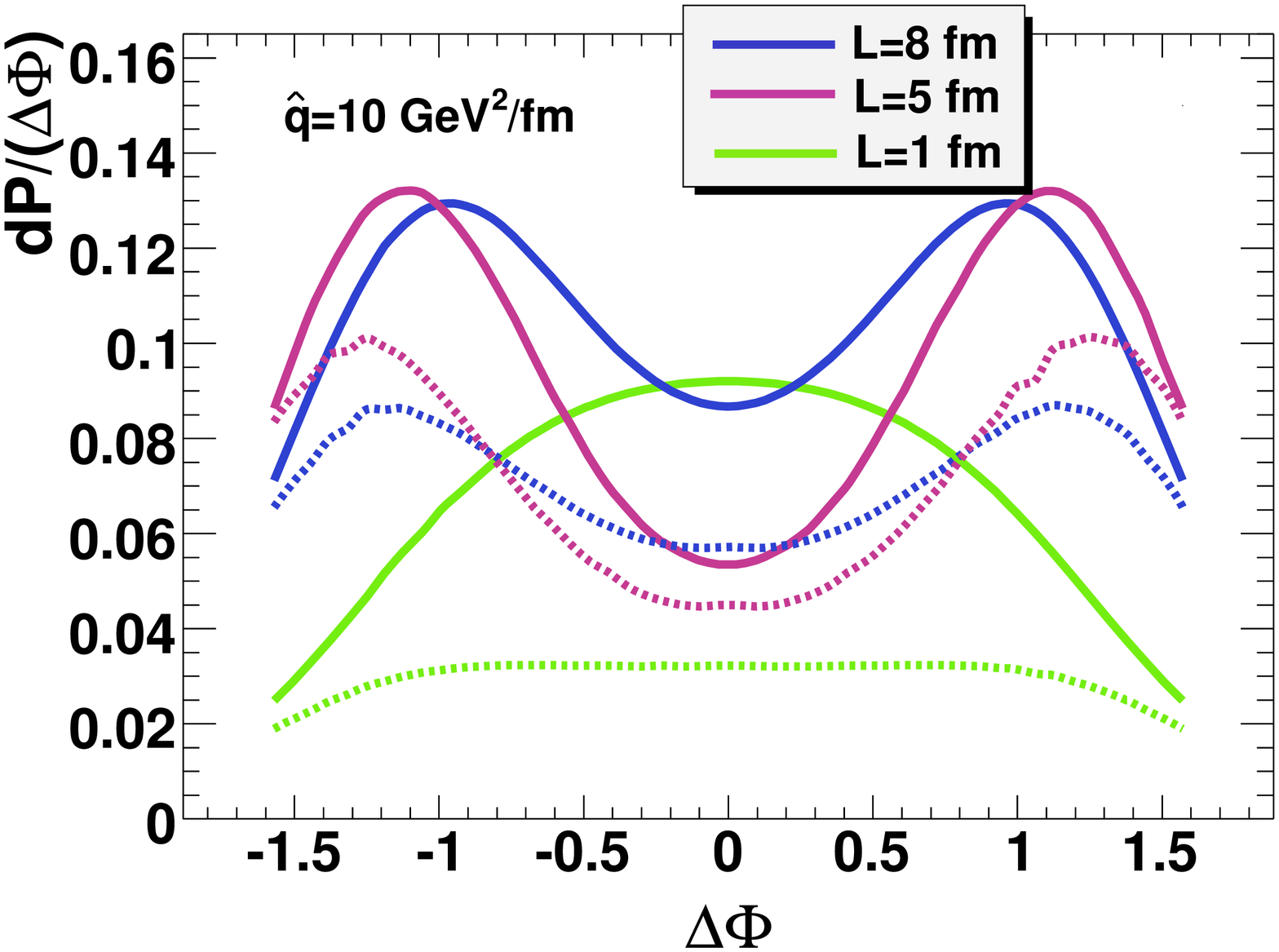}
\end{center}
\end{minipage}
\hfill
\begin{minipage}{0.5\textwidth}
\begin{center}
\includegraphics[width=0.75\textwidth,angle=-90]{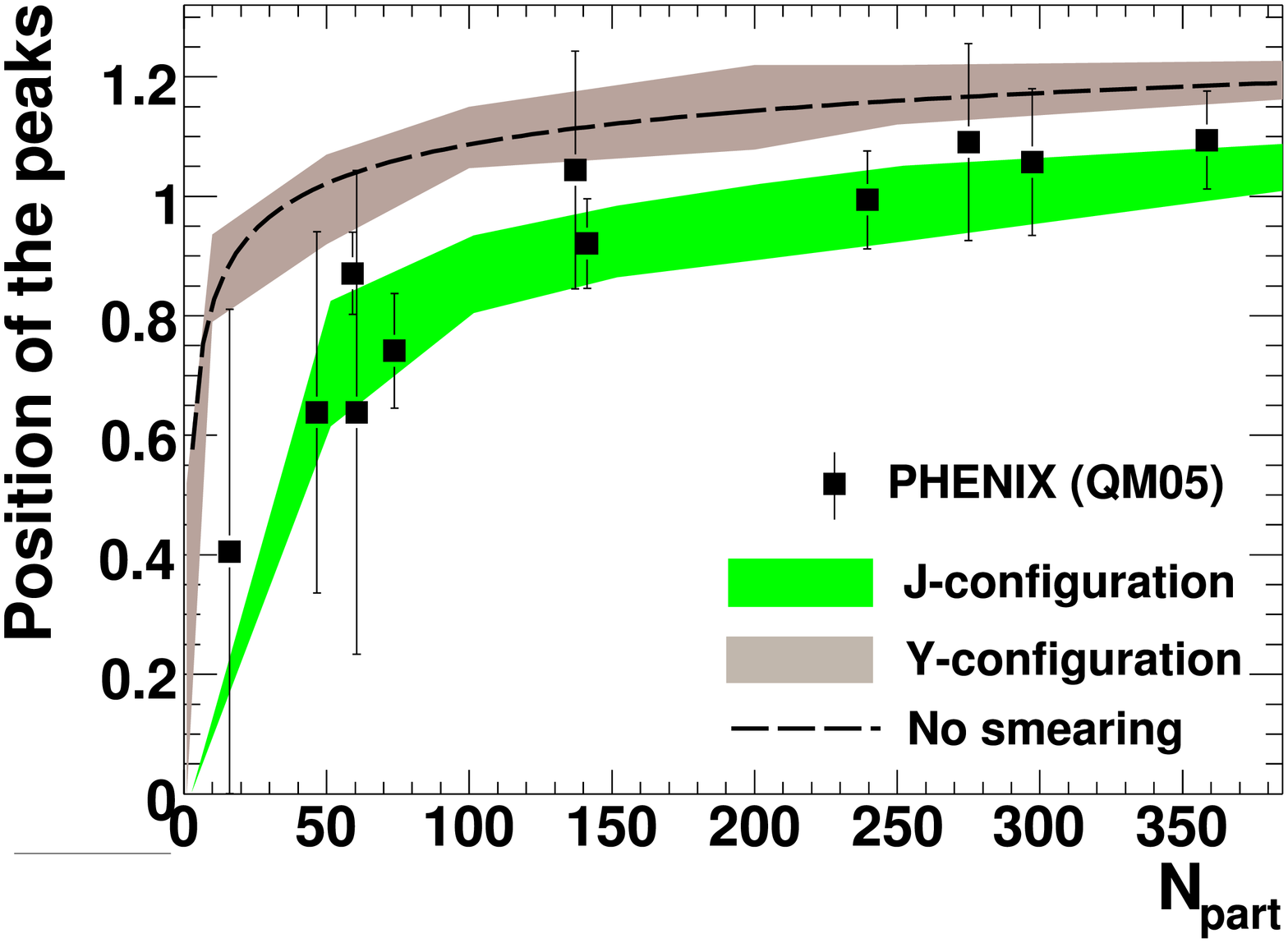}
\end{center}
\end{minipage}
\caption{Left: The probability of just one splitting (\protect\ref{eq:splitlab}) \cite{Polosa:2006hb} as a function of the laboratory azimuthal angle $\Delta\Phi$ for a gluon jet of $E_{\rm  jet}=7$ GeV. Right: Position of the peaks of the $\Delta\Phi$-distribution and comparison with PHENIX data from Ref. \cite{Grau:2005sm}.}
\label{fig:ndist}
\end{figure}

The inclusive measurements study only one of the medium effects on high-$p_t$ partons traversing a medium, namely, the additional amount of energy loss. This energy loss, if dominated by radiated mechanisms as our results seem to favor, is accompanied by a characteristic change in the associated jet properties. These type of studies of heavy flavor initiated jets are almost inexistent for heavy--ion collisions -- see \cite{Antinori:2005tu} for related a proposal. If experimental measurements are possible, it could provide very valuable information about the dynamics, as it is the case for light partons. Let us comment here on how such a calculations could be performed. For the case of light quarks, a recent proposal \cite{Polosa:2006hb} relate the surprising two-peak shapes in the away-side azimuthal correlation distribution \cite{Adler:2005ee}
 with the exclusive one-gluon splitting in the radiative energy loss formalism. More explicitly, the probability of splitting for radiated gluons with energy $\omega\lesssim (\hat q)^{1/3}$ [$\sim$ 3 GeV for central AuAu collisions at RHIC]
\begin{equation}
\frac{d{ P}(\Phi,z)}{dz\,d\Phi}\Bigg\vert_{\eta=0}=
\frac{\alpha_s C_R}{16\pi^2}\,E\,L\,\cos\Phi
\exp\left\{-E\,L\,\frac{\alpha_s C_R}{16\pi}\cos^2\Phi\right\}
\label{eq:splitlab}
\end{equation}
present well defined peaks in the laboratory azimuthal angle $\Phi$. Although the analysis is oversimplified, it is interesting to see that the obtained shape resembles very much experimental data -- see Fig. \ref{fig:ndist}. This peaks should disappear for larger values of $\omega\sim p_t^{\rm assoc}$.

Mass effects make the medium--induced gluon radiation more collinear than the corresponding one for light quarks -- see Fig. \ref{fig:spec}. This is in contrast to the vacuum case where exactly the opposite happens. Moreover, the suppression effect of the Sudakov form factor -- the exponent factor in eq. (\ref{eq:splitlab}) -- is generally smaller due to the smaller integrated radiation. So, qualitatively, the two-peak structure, if present at all in the case of heavy-quark jets, should be more collinear than in the light parton case. 

\section{Final comments}

We have presented comparisons with available experimental data and predictions for heavy quark production in nuclear collisions. Our results provide a reasonable description of non-photonic electrons data in central AuAu collisions at RHIC, in the presence of important benchmark uncertainties. These results are a direct prediction from the formalism once the transport coefficient is fixed from other data sets. Improvements in the formalism are possible, e.g. following the lines sketched in section \ref{sec:evol}. However, a definite answer for this important subject would only be possible with a better control on the benchmark, ideally by direct measurements of $D$ and $B$ mesons.

\noindent{\bf Acknowledgements}\\
This work is supported by the 6th Framework Programme of the European Community under the Marie Curie contract MEIF-CT-2005-024624.

\end{document}